\documentclass[prl,twocolumn,superscriptaddress,noshowpacs]{revtex4}
\usepackage{graphicx}
\usepackage{physics}
\usepackage{bm}
\usepackage{ulem}
\usepackage{color}

\begin{document}

\title{The Berry-Foucault Pendulum}

\author{D.~D.~Solnyshkov}
\affiliation{Institut Pascal, PHOTON-N2, Universit\'e Clermont Auvergne, CNRS, Clermont INP, F-63000 Clermont-Ferrand, France.}
\affiliation{Institut Universitaire de France (IUF), F-75231 Paris, France}
\author{I.~Septembre}
\affiliation{Institut Pascal, PHOTON-N2, Universit\'e Clermont Auvergne, CNRS, Clermont INP, F-63000 Clermont-Ferrand, France.}
\author{K.~Ndiaye}
\affiliation{Institut Pascal, PHOTON-N2, Universit\'e Clermont Auvergne, CNRS, Clermont INP, F-63000 Clermont-Ferrand, France.}
\author{G.~Malpuech}
\affiliation{Institut Pascal, PHOTON-N2, Universit\'e Clermont Auvergne, CNRS, Clermont INP, F-63000 Clermont-Ferrand, France.}

\begin{abstract}
The geometric phase is known to play a role both in the rotation of the Foucault pendulum and in the anomalous Hall effect (AHE) due to the Berry curvature. Here, we show that a 2D harmonic oscillator with AHE induced by Berry curvature behaves exactly like the Foucault pendulum: in both, the plane of the oscillations rotates with time. The rotating pendulum configuration enhances the AHE, simplifying its observation and allowing high-precision measurements of the Berry curvature. We also show how the non-adiabaticity and anharmonicity determine the maximal rotation angle and find the optimal conditions for the observations.
\end{abstract}


\maketitle

The Foucault pendulum was suggested and implemented by Léon Foucault in 1851 to demonstrate the Earth's rotation \cite{Foucault1851}. It is one of the most widely known experiments in popular culture \cite{conlin1999popular}, demonstrated in many science museums around the world \cite{oprea1995geometry}. However, it is much less known that it was also Léon Foucault who, inspired by his pendulum's working principle, created the first gyroscope \cite{Gilbert1882,sommeria2017foucault}. Contrary to the Foucault pendulum, which mostly remains a brilliant demonstration, the gyroscopes, based on the same basic property, have become extremely widespread, starting from the navigation applications. Now they are used for the orientation at all scales, from smartphones~\cite{Lane2010} to the International Space Station~\cite{Bedrossian2009}. Moreover, some of the most efficient microscopic implementations of the gyroscopes \cite{Yazdi1998} use  oscillations instead of rotation \cite{Shao2014}, thus coming back to the Foucault pendulum (on a chip) \cite{Prikhodko2012}. Recently, an implementation of the Foucault pendulum based on a Bose-Einstein condensate in a synthetic rotational field has been suggested \cite{Qu2018}. 

Geometric phases are quite widespread in physics \cite{GeomPhases}. Such phase plays a key role in the Foucault pendulum by determining its rotation versus latitude \cite{Hannay1985,Berry1985class,Khein1993,oprea1995geometry,vonBergmann2007,Delplace2020}. The geometric phase is accumulated during the transport around the Earth at a given latitude. The associated rotation angle can be linked with an integral of the Earth's curvature.  The Foucault pendulum can also be used to measure the Earth's gravitomagnetic field~\cite{Braginsky1984} or the Lense-Thirring effect \cite{cartmell2020modelling}. From an even broader perspective, since any non-inertial motion can be viewed as a spacetime curvature in general relativity, any rotation of the Foucault pendulum is always due to a certain curvature (Earth's curvature, spacetime curvature or some other), and thus the Foucault pendulum can be seen as a device for the measurement of curvatures and of the associated geometric phases.

The Berry (Pancharatnam) phase is another well-known example of a geometric phase, studied in quantum \cite{Pan1956,berry1984quantal,berry1989quantum} and classical systems~\cite{Berry1985class}: it is the curvature of the parameter space of the Hamiltonian, for example, the reciprocal space. This phase and the associated Berry curvature are key concepts in modern topological physics, being involved in the optical spin Hall effect \cite{Onoda2004,Bliokh2004,Kavokin2005,Bliokh2008}, topological insulators \cite{Konig2007,hsieh2008topological,Hasan2010}, lasers \cite{Bahari2017}, and optical isolators \cite{Solnyshkov2018,Karki2019}. The Berry curvature can be seen as a gauge field, an equivalent of a magnetic field in the reciprocal space. As such, it can affect the spatial trajectories of the particles, leading to the Anomalous Hall Effect (AHE) \cite{Sundaram1999}, recently measured in a photonic system \cite{gianfrate2020measurement}. 
Two general 2D Hamiltonians characterized by a non-zero Berry curvature are the Dirac Hamiltonian (winding 1) and the TE-TM Hamiltonian (winding 2). The former is  the most well-known Hamiltonian exhibiting Berry curvature and representing a reference case. Its implementations include 2D electron gas with Rashba Spin Orbit Coupling (SOC) \cite{bychkov1984properties} and Zeeman splitting \cite{Bolte2007,Zawadzki2011}, 2D transitional metal dichalcogenides \cite{Xiao2012,Wang2018}, biased bilayer graphene \cite{ju2015topological}, and  photonic quantum valley Hall effect \cite{Noh2018}.
The TE-TM Hamiltonian describes the inherent topology of the photonic modes associated with their vectorial nature \cite{bliokh2008geometrodynamics}. It describes the propagation of light beams in any inhomogeneous system, for example, for paraxial beams of light \cite{bliokh2008geometrodynamics,jisha2017paraxial,Zhang2020} and in microcavities~\cite{Kavokin2005,leyder2007observation}. In recent studies it appears combined with non-Hermiticity \cite{su2021direct,Krol2022,Hu2022}. Photonic systems are interesting because they allow direct access to the dispersion of modes, their Berry curvature distribution, and real space dynamics of wave packets in various potential profiles. With linear birefringence, they also allow implementing the 2D massive Dirac Hamiltonian~\cite{Tercas2014,gianfrate2020measurement,polimeno2021tuning}.

Inspired by the generality of the geometric phase, in this work we study a 2D harmonic oscillator with Berry curvature. The accumulation of the associated geometric (Berry) phase leads to the rotation of the plane of the oscillations, as in the Foucault pendulum, but without any non-inertiality. We call this system the Berry-Foucault pendulum. We consider two above-mentioned Hamiltonians (Dirac and TE-TM) in two dimensions, showing how the AHE can be amplified in the Foucault pendulum configuration, allowing to measure the Berry curvature with a high precision and even in systems, where the eigenstates are not directly accessible. We also study the limits of the Berry-Foucault pendulum operation set by the non-adiabaticity and the anharmonicity, determining the optimal configuration for measurements.

\begin{figure}[tbp]
    \centering
    \includegraphics[width=0.99\linewidth]{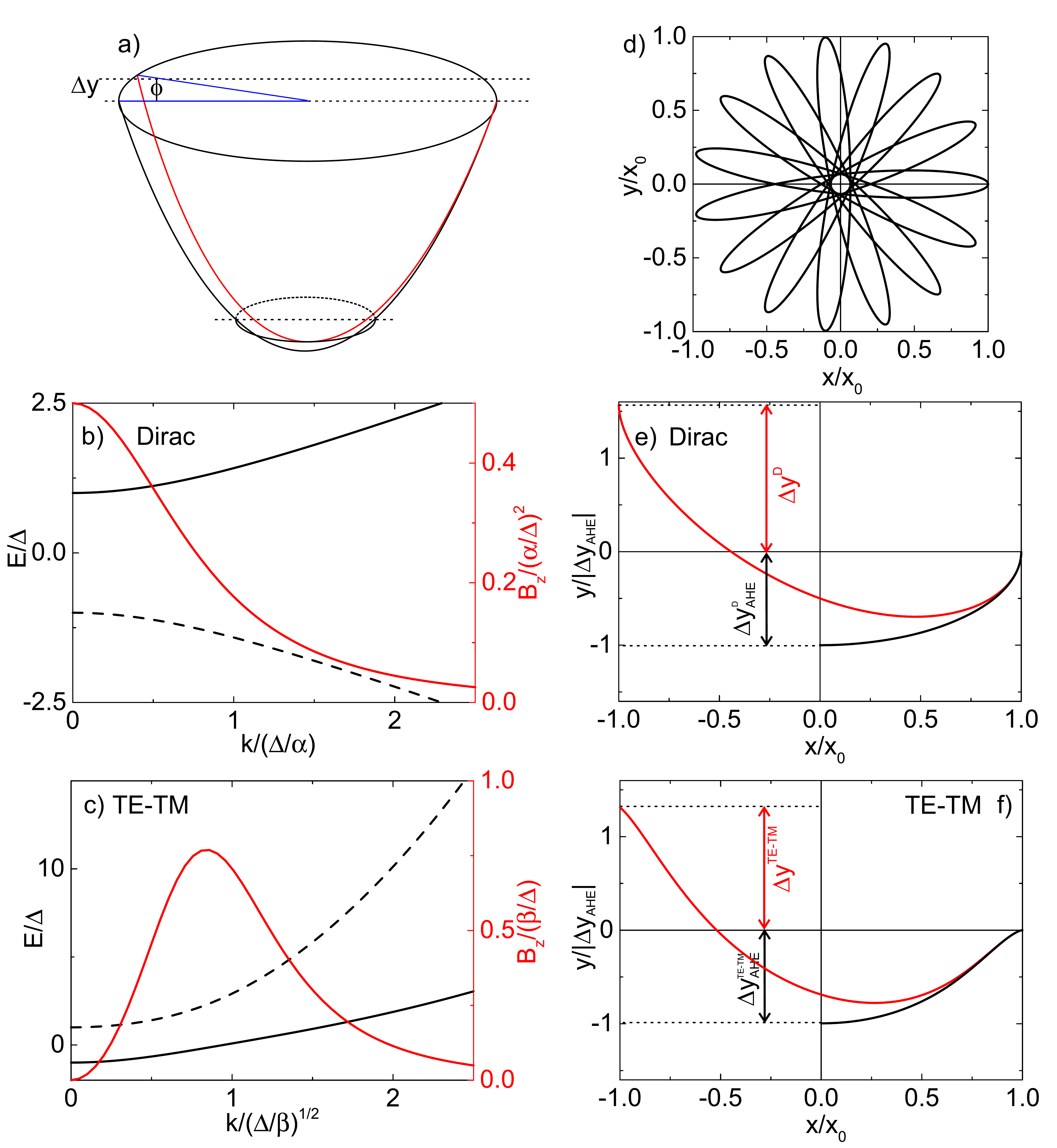}
    \caption{(a) Parabolic potential and wavepacket trajectory over half oscillation period with a deviation $\Delta y$ corresponding to an angular shift of $\phi$. (b) Trajectory of a Berry-Foucault pendulum in the case of a Dirac Hamiltonian~\eqref{DirHam}, completely equivalent to a Foucault pendulum (reduced units). (c,d) Energy spectrum (black) and Berry curvature (red) profile for the Dirac Hamiltonian~\eqref{DirHam} (c) and the TE-TM~\eqref{TETMHam} Hamiltonian (d). Black solid lines are the branches of interest. (e,f) Comparison between the AHE $\Delta y_{AHE}$ (black) and the Berry-Foucault deviation $\Delta y$ (red) for Dirac (e) and TE-TM (f).}
    \label{fig1}
\end{figure}

\emph{The model.}
We consider a wavepacket oscillating in a harmonic oscillator potential $U(x,y)=\xi(x^2+y^2)/2$  (Fig.~\ref{fig1}(a)). A wavepacket is launched at $x=x_0$, and the accumulation of the Berry phase leads to the AHE and thus to the spatial deviation $\Delta y$ of the trajectory, leading to the rotation $\phi$ of the oscillation plane. 

The system is described either by the Dirac, or by the TE-TM Hamiltonian.
The Dirac Hamiltonian reads:
\begin{equation}
\hat H_{Dirac} = \left( {\begin{array}{*{20}{c}}
{ + \Delta_D }&{\alpha k{e^{ - i\varphi }}}\\
{\alpha k{e^{i\varphi }}}&{ - \Delta_D }
\end{array}} \right)
\label{DirHam}
\end{equation}
where $k$ is the 2D wavevector modulus, $\varphi$ its polar angle. In the original Dirac equation, $\Delta_D=m_e c^2$, $\alpha=\hbar c$, where $m_e$ is the electron mass.
The Berry curvature of the lower band reads:
\begin{equation}
    B_z^D\left(k\right)=\frac{\alpha^2 \Delta_D}{2\left(\alpha^2 k^2+\Delta_D^2\right)^{3/2}}
    \label{DirBer}
\end{equation}

The TE-TM Hamiltonian with a time-reversal symmetry breaking term $\Delta$ (Faraday effect) reads: 
\begin{equation}
\hat H_{TE-TM} = \left( {\begin{array}{*{20}{c}}
{\frac{{{\hbar ^2}{k^2}}}{{2m}} + \Delta }&{\beta {k^2}{e^{ - 2i\varphi }}}\\
{\beta {k^2}{e^{2i\varphi }}}&{\frac{{{\hbar ^2}{k^2}}}{{2m}} - \Delta }
\end{array}} \right)
\label{TETMHam}
\end{equation}
The Berry curvature of the lower band is
\begin{equation}
    B_z^{TE-TM}\left(k\right)=\frac{2\Delta \beta^2 k^2}{\left(\Delta^2+\beta^2 k^4\right)^{3/2}}
    \label{TETMBer}
\end{equation}

Figure~\ref{fig1}(b,c) shows the two band dispersions and the lower band Berry curvature for the two Hamiltonians. The Dirac Hamiltonian is well described by a parabolic dispersion at $k\approx 0$ and exhibits the maximum of the Berry curvature at the same point, while the TE-TM Hamiltonian has the maximum of the Berry curvature at $k=\sqrt{\Delta/\sqrt{2}\beta}\neq 0$, where the dispersion is non-parabolic. 

\emph{Semiclassical description.}
The qualitative understanding of the AHE came with the semiclassical equations derived by Sundaram and Niu \cite{Sundaram1999}: 
\begin{eqnarray}
\Dot{\bm{p}}&=&-\nabla U\\
\Dot{\bm{r}}&=&\nabla_p E+\hbar^{-1}\Dot{\bm{p}}\times \bm{B}(\bm{p})
\label{eqSC}
\end{eqnarray}
where the Berry curvature appears as a kind of magnetic field acting in the reciprocal space. The analogy between the AHE and the Lorentz force is not full in the general case.  Nevertheless, for a harmonic oscillator and a constant Berry curvature (Dirac Hamiltonian,  $k\approx 0$) that can be considered as a correction, it is possible to obtain a complete reciprocity between the two effects (see ~\cite{suppl}, sec. I for details).  There is also a complete analogy between the magnetic part of the Lorentz force the non-inertial Coriolis force acting on the Foucault pendulum. This makes the Berry-Foucault pendulum with the Dirac Hamiltonian fully mathematically equivalent to the Foucault pendulum. The corresponding trajectories, identical in both cases (Foucault and Berry-Foucault), are shown in Fig.~\ref{fig1}(d). The TE-TM case with a variable Berry curvature shows, of course, a different behavior.

The next step is to quantitatively compare the classical AHE, where a 1D accelerating potential provokes a transverse deviation, with the 2D harmonic oscillator case. We take the initial condition $x_0>0$, $y_0=0$, $\bm{p}_0=0$. In the AHE case, the deviation is given by
\begin{equation}
    \Delta y_{AHE}=\int\limits_0^{k_{max}} B(k)dk
    \label{yAHE}
\end{equation}
where $k_{max}$ is the maximal wavevector achieved during the acceleration. In the Dirac case it reads: 
\begin{equation}
    \Delta y_{AHE}^{D}=\frac{\alpha^2 k_{max}}{2\Delta_D\sqrt{\alpha^2k_{max}^2+\Delta_D^2}}
\end{equation}
It grows with $k_{max}$ up to a certain limit determining the maximal deviation.
For small $k_{max}$ it can be rewritten as 
$\Delta y_{AHE}^{D}=B(0) k_{max}$, 
where $B(0)=\alpha^2/2\Delta_D^2$ is the Berry curvature at $k=0$.
For the TE-TM case, the AHE drift is given by the hypergeometric  function: 
 \begin{equation}
    \Delta y_{AHE}^{TE-TM}=\frac{2\beta^2k_{max}^3~{}_2F_1(1/4,1,7/4,-\beta^2k_{max}^4/\Delta^2)}{3\Delta\sqrt{\Delta^2+\beta^2k_{max}^4}}
\end{equation}
which for small $k_{max}$ reads:
\begin{equation}
    \Delta y_{AHE}^{TE-TM}\approx \frac{2\beta^2 k_{max}^3}{3\Delta^2}
\end{equation}

We then consider the harmonic oscillator case assuming $\Delta y\ll x_0$, which allows to decouple $x$ and $y$. In that case, the equations ~\eqref{eqSC} for $y(t)$ and $p_y(t)$ become equivalent to a driven harmonic oscillator with an external force defined by $p_x B_z(p_x)$. 
Using the Green's function approach (\cite{suppl}, I), we find the ratios of the deviations 
$\Delta y^D/\Delta y_{AHE}^D=-\pi/2$ and  $\Delta y^{TE-TM}/\Delta y_{AHE}^{TE-TM}=-3\pi/8$. In both cases, the deviation for a half-period is comparable in magnitude, but has a direction \emph{opposite} with respect to the AHE deviation. The AHE and a half-period Berry-Foucault pendulum trajectories are shown in Fig.~\ref{fig1}(e,f). At early time both coincide, but then, the 2D parabolic potential provides its own lateral acceleration, bringing a lateral deviation opposite to the initial one of the AHE. The same occurs for the original Foucault pendulum, whose trajectory is exactly similar to the one of the Dirac case.

For small wave vectors, $k_{max}$ in the harmonic oscillator can be obtained from $x_0$ (the oscillation amplitude) as: $k_{max}^D\approx x_0\sqrt{\Delta_D\xi}/\alpha$ and $k_{max}^{TE-TM}\approx x_0 \sqrt{\xi m}/\hbar$. Thus, the angles of rotation of the oscillation plane for a half-period are
\begin{equation}
\label{phiD}
    \phi^D=\arctan\frac{\Delta_{y}^D}{x_0}\approx\frac{\pi\alpha\sqrt{\xi}}{4\Delta_D^{3/2}},
\end{equation}
which in this limit does not depend on $x_0$, and
\begin{equation}
\label{phiTETM}
    \phi^{TE-TM}\approx \frac{\pi\beta^2 \xi^{3/2} m^{3/2} x_{0}^2}{4\hbar^3\Delta^2},
\end{equation}
which grows quadratically with $x_0$.

To stress that the Foucault pendulum can be seen as an instrument for curvature measurements, we compare the Dirac case with the well-known Earth's curvature measurement provided by the rotation of the Foucault pendulum (see \cite{suppl}, II for details). The similarity of the two expressions is due to the geometric origin of the effect (anholonomy).

\begin{figure}[tbp]
    \centering
    \includegraphics[width=0.99\linewidth]{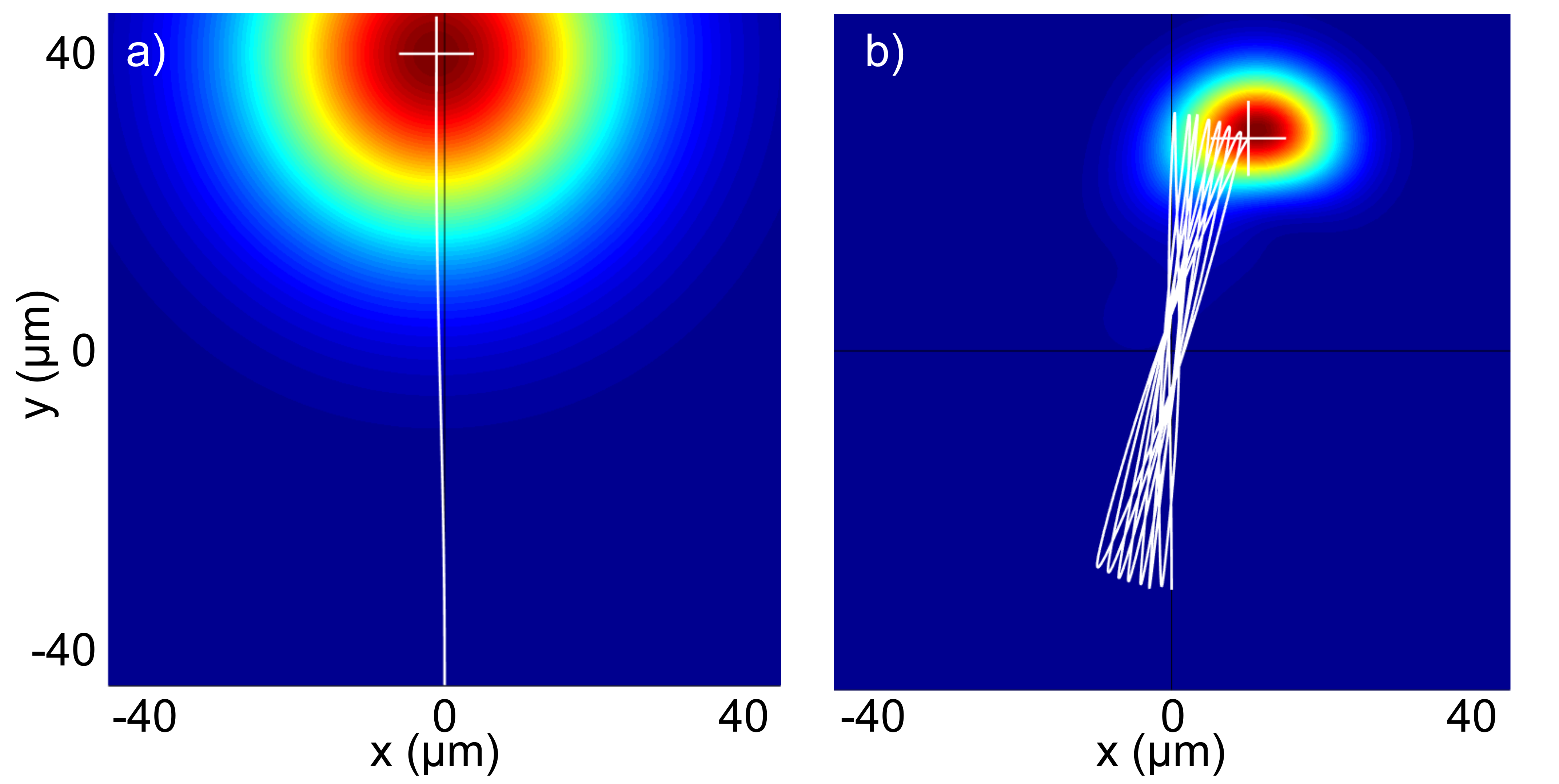}
    \caption{Amplification of the Berry curvature effect.  a) Reference: AHE; b) Amplification: Berry-Foucault pendulum. False color -- particle density $n=|\psi|^2$, white line -- center of mass trajectory.}
    \label{fig2}
\end{figure}

\emph{Numerical simulations.}
The next step is to go beyond the semi-classical picture considering that we are dealing with spatially extended wave packets and not classical particles. For this, we perform numerical simulations with a time-dependent Schrödinger equation $i\hbar\partial\psi/\partial t=\hat{H}\psi$. Fig.~\ref{fig2} shows these simulations for the TE-TM Hamiltonian \eqref{TETMHam} for the AHE (panel a) and the Berry-Foucault pendulum (panel b). The parameters used\cite{params2} are typical for GaAs-based microcavities ~\cite{klembt2018exciton}.
The AHE, resulting from a constant potential gradient, requires working with a large wavepacket: the scale of the effect is much smaller than the wavepacket size $\Delta y\ll\sigma$. Even if a smaller wavepacket is taken initially, it necessarily expands over time. For realistic parameters the AHE appears as a slight drift with respect to the straight trajectory (white line). 

The Berry-Foucault pendulum (Fig.~\ref{fig2}(b)) is based on the use of a 2D parabolic potential, whose parameters determine the size of the wavepacket: $\sigma=\sqrt{\hbar/m\omega}=\sqrt{\hbar/\sqrt{\xi m}}$. 
Increasing the stiffness $\xi$  decreases the wavepacket size. In that case, the wavepacket does not expand or shrink over time. This allows to operate with wavepackets much smaller than for the AHE, as immediately visible in Fig.~\ref{fig2}(b). Moreover, the Berry-Foucault pendulum carries out multiple oscillations (up to $Q$), and thus the anomalous Hall deviation is amplified accordingly, allowing the final deviation to exceed the wavepacket size: $Q\Delta y>\sigma$. 

\begin{figure}[tbp]
    \centering
    \includegraphics[width=0.99\linewidth]{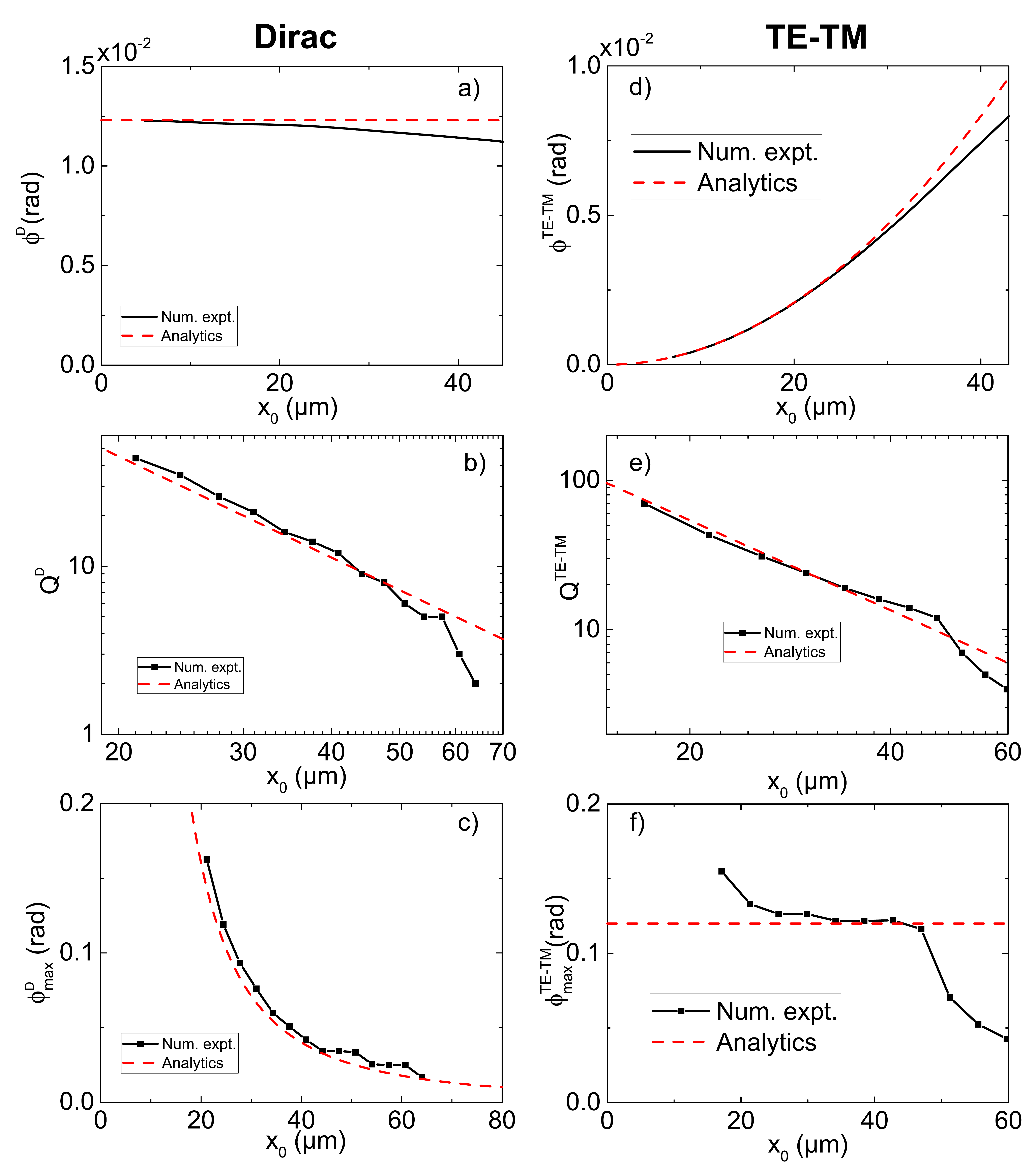}
    \caption{Berry-Foucault pendulum for the Dirac and TE-TM Hamiltonians respectively:  (a,d) half-period rotation angle $\phi$; (b,e) Q-factor $Q$ as a function of initial position $x_0$; (c,f) maximal rotation angle $\phi$.  Black lines and points - numerical experiment, red dashed lines - analytics.}
    \label{fig3}
\end{figure}

We then compare the analytical results for the rotation angle \eqref{phiD},\eqref{phiTETM} with those of Schrödinger equations for a half-period (see \cite{params} for parameters). Figure~\ref{fig3}(a,b) shows the $x_0$ dependence. The Schrödinger equation  reproduces the limit of $\phi^D$ for $x_0\to 0$. The quadratic dependence of $\phi^{TE-TM}$ is well reproduced at low $x_0$ as well, before to deviate for higher values.


\emph{Limitations.} The maximal number of oscillations is determined by the Q-factor of the oscillator $Q$, controlling the maximal rotation angle $\phi_{max}$. The losses are determined by the non-adiabaticity and by the anharmonicity. Non-adiabaticity is related to the transfer of the wave packet from the initial band to the other band because of the finite acceleration.  Anharmonicity is induced by the non-parabolic character of the dispersion. Similar limitations exist also for the classical Foucault pendulum, which deviates from perfectly planar oscillations after 1~hour of operation~\cite{Schulz1970}, unless suppression devices are used~\cite{sommeria2017foucault}. From the experimental point of view, the simplest tunable variable is $x_0$. 

A convenient tool to estimate the non-adiabaticity when a wave packet propagates in a band is the quantum metric $g_{\bm{kk}}$ \cite{Bleu2018effective}. The non-adiabatic fraction can be estimated as $\sqrt{f_{NA}}\approx \sqrt{g_{kk}}k_{max}$. 
For the Dirac Hamiltonian at $k\approx 0$, $g_{kk}=\alpha^2/4\Delta_D^2$ (see \cite{suppl}, III). Using the expression for $k_{max}$, we obtain $f_{NA}^D\approx \xi x_0^2/4\Delta_D$. At each half-period, the non-adiabatic fraction escapes the confinement due to the opposite sign of the associated mass. The non-adiabaticity thus directly determines the losses (see \cite{suppl}, III for anharmonicity). This allows to determine the Q-factor of the oscillator as:
\begin{equation}
    Q^D=\frac{1}{f_{NA}^D}=\frac{4\Delta_D}{\xi x_0^2}.
    \label{finesseDirac}
\end{equation}
 Finally, the maximal rotation angle is
\begin{equation}
    \phi_{max}^D=\frac{\phi_D}{f_{NA}^D}=\frac{\pi\alpha}{\sqrt{\xi\Delta_D}x_0^2}
\end{equation}

For the TE-TM Hamitonian, $g_{kk}=\beta^2k^2/\Delta^2$. The non-adiabatic fraction turns out to be negligible with respect the anharmonicity (\cite{suppl}, III). The non-parabolicity of dispersions makes the states of the harmonic potential unequally spaced in frequency. The oscillating wave packet is a linear superposition of these eigenstates. For small $x_0$, anharmonicity simply provokes a decay of the wavepacket, allowing to write perturbatively: 
$Q=\omega/\Delta\omega$, where $\omega=\sqrt{\xi/\bar{m}}$ is the mean frequency of the harmonic oscillator and $\Delta\omega$ is due to the mass difference between $k=0$ and $k=k_{max}$.  Beyond a critical $x_0$, the deviation from the ideal harmonic oscillator quickly destroys the wave packet~(\cite{suppl}, IV). 

These analytical predictions are compared with the results of numerical experiments in Fig.~\ref{fig3}(b,c), with all panels demonstrating a good agreement. The Q-factor $Q(x_0)$ is plotted in panel (b) in double log scale, demonstrating the power law decay up to critical $x_0$, as predicted by~\eqref{finesseDirac}. Accordingly, the maximal rotation angle also decays with $x_0$ (Fig.~\ref{fig3}~(c)).

For the TE-TM Hamiltonian, $m(k)\approx m(1+6\beta^2 k^2 m/\hbar^2 \Delta)$,
which gives
\begin{equation}
    Q^{TE-TM}\approx \frac{\hbar^4 \Delta}{6\beta^2 m^2 \xi x_0^2}
\end{equation}
This allows obtaining the maximal rotation angle:
\begin{equation}
    \phi_{max}^{TE-TM}\approx \frac{\pi\xi^{1/2}}{12\Delta m^{1/2}}=\frac{\pi}{12}\frac{\hbar\omega}{\Delta}
\end{equation}
The maximal rotation angle therefore does not depend on the initial position $x_0$, contrary to the Dirac case, where it was decaying with $x_0$. Moreover, it does not depend on one of the parameters of the Hamiltonian, the TE-TM splitting $\beta$, because the dependencies of $\phi$ and $Q$ on $\beta$ are opposite and compensate each other.

The Q-factor $Q^{TE-TM}$ versus $x_0$ is plotted in Fig.~\ref{fig3}(e) and compared with results of the Schrödinger equation. It shows an excellent agreement up to a critical $x_0$ value, where $Q$ quickly drops. With the realistic parameters $Q$ can reach values as high as $10^2$. The total rotation angle $\phi_{max}^{TE-TM}$ is plotted in Fig.~\ref{fig3}(f) as a function of $x_0$, showing a good agreement with the analytical estimate for small $x_0$ up to the same critical cut-off $x_0$ value (see \cite{suppl}, IV).

Another important parameter for a practical experiment in systems with finite lifetime $\tau$ is the frequency of rotation of the plane of oscillations of the Berry-Foucault pendulum which can be found as $\Omega=2\phi\omega$. Within the same limit as before, it gives $\Omega\sim \pi\beta^2\xi^2 m x_0^2/2\hbar^3\Delta^2$. One could expect  $\tau\gg\Omega^{-1}$ for the experiment to be carried out. However, the polaritonic wavepackets can be reamplified from excitonic reservoir \cite{Wertz2012}, similar to the classical Foucault pendulums accelerated electromagnetically in museums~\cite{sommeria2017foucault}. We therefore consider the condition $\tau\gg \Omega^{-1}$ to be less stringent than the other limitations discussed above.

The Berry-Foucault pendulum allows to improve the precision of the measurement of the Berry curvature by the factor $Q$ (see \cite{suppl}, V). This would mean reducing the $10\%$ uncertainty on the Berry curvature in Ref.~\cite{gianfrate2020measurement} down to $0.1\%$ with $Q=100$ from Fig.~\ref{fig3}, a significant improvement. Its limit is set by $\Delta>\beta/\sigma^2=4\pi^2\beta\sqrt{\xi m}/\hbar$ (wavepacket size should be smaller than the Berry curvature variation scale).

To conclude, we have demonstrated that the Berry curvature leads to the Foucault pendulum effect for a harmonic oscillator with spin-orbit coupling. We have studied two most well-known Hamiltonians and determined the Q-factor and the maximal rotation angle. The Berry-Foucault pendulum can be used for high-precision curvature measurements close to the band extrema for systems, where the direct measurement of the eigenstates is impossible (such as electronic systems).

\begin{acknowledgments}
We acknowledge the support of the European Union's Horizon 2020 program, through a FET Open research and innovation action under the grant agreement No. 964770 (TopoLight), project ANR Labex GaNEXT (ANR-11-LABX-0014), and of the ANR program "Investissements d'Avenir" through the IDEX-ISITE initiative 16-IDEX-0001 (CAP 20-25). 
\end{acknowledgments}

\bibliography{biblio}

\begin{thebibliography}{58}
\bibitem[{par({\natexlab{a}})}]{params2}
\bibinfo{note}{The parameters used in numerical simulations for Fig.~2 were:
  the polariton mass $m=5\times10^{-5}m_0$ ($m_0$ is the free electron mass),
  polariton lifetime $\tau=300$~ps, $\Delta=0.1$~meV,
  $\beta=270~\mu$eV$\mu$m$^2$, $\xi=1.6\times 10^{-13}$~N/m. The simulations
  were carried out using the 3rd-order Adams method for the time derivative and
  a GPU-accelerated FFT for the Laplacian.}
\bibitem[{par({\natexlab{b}})}]{params}
\bibinfo{note}{The parameters used in numerical simulations for Fig.~3 were,
  for the Dirac Hamiltonian: $\alpha=2$~meV$\mu$m, $\Delta_D=2$~meV,
  $\xi=7.8e-14$~N/m. For the TE-TM Hamiltonian they were: the polariton mass
  $m=2\times10^{-5}m_0$ ($m_0$ is the free electron mass), polariton lifetime
  $\tau=300$~ps, $\Delta=0.24$~meV, $\beta=135~\mu$eV$\mu$m$^2$, $\xi=4.4\times
  10^{-13}$~N/m. The simulations were carried out using the 3rd-order Adams
  method for the time derivative and a GPU-accelerated FFT for the Laplacian.}
\end{thebibliography}

\section{Supplemental materials}

\renewcommand{\thefigure}{S\arabic{figure}}
\setcounter{figure}{0}
\renewcommand{\theequation}{S\arabic{equation}}
\setcounter{equation}{0}

In these Supplemental Materials, we present the details concerning the Berry-Foucault pendulum: the description of the anomalous Hall deviation, a comparison with the measurement of the Earth's curvature, the non-adiabaticity and anharmonicity, the limitations of the perturbative treatment, the uncertainty, the determination of the full Berry curvature profile, and the large-amplitude case.

\subsection{The anomalous Hall effect in the Berry-Foucault pendulum configuration}

First, we note that there is a complete analogy between the non-inertial Coriolis force and the magnetic part of the Lorentz force. However, the analogy between the anomalous Hall effect and the Lorentz force is not full in the general case. 
 
A magnetic field can be incorporated into the Hamiltonian as a vector potential $\bm{A}$: $H_{\mathrm{mag}}=(\bm{p}-e\bm{A})^2/2m$ containing the term $\bm{p}\cdot\bm{A}$, whose reciprocal term should be $\bm{r}\cdot\bm{A}$,
whereas the Berry curvature would correspond to a term $\Dot{\bm{p}}\cdot\bm{A}$ instead. This shows up in the fundamental difference of trajectories between the ordinary Hall effect (circles in 2D) and the AHE (finite deviation of the curve).

Nevertheless, for a harmonic oscillator and a constant Berry curvature (Dirac Hamiltonian, $k\approx 0$) that can be considered as a correction, it is possible to obtain complete reciprocity between the two effects
by using the unperturbed solution, for which $\Dot{\bm{p}}\approx-\bm{r}$.  The symmetry of Hamilton's equations of motion concerning the substitution $\bm{r}'=-\bm{p}$, $\bm{p}'=\bm{r}$ allows establishing a full analogy between the ordinary Foucault pendulum and the Berry-Foucault pendulum in the Dirac case.

We then consider the harmonic oscillator case assuming $\Delta y\ll x_0$ which allows to decouple $x$ and $y$, considering $x(t)=x_0\cos\omega t$ and $p_x(t)=-p_{\mathrm{max}}\sin\omega t$ as known. This allows us to write a system of equations for $y(t)$ and $p_y(t)$ from~\eqref{eqSC}. 
\begin{eqnarray}
\Dot{p}_y&=&-\xi y\\
\Dot{y}&=&p_y/m-\hbar^{-1}\Dot{p}_x B_z(p_x)
\label{eqSCs}
\end{eqnarray}
This system of equations is equivalent to a driven harmonic oscillator with an external force defined by $p_x B_z(p_x)$.

Its solution is given by
\begin{equation}
    \Delta y=-p_{\mathrm{max}}\omega^2\int\limits_0^{T/2} \cos^2\omega t' B(p_{\mathrm{max}}\sin\omega t')\,dt'
\end{equation}
with $T=2\pi/\omega$. The integration gives the coefficients $-\pi/2$ and $-3\pi/8$ given in the main text.

\subsection{Foucault pendulum and Earth's curvature}

In the main text, we have obtained the angle of rotation of the Berry-Foucault pendulum with the Dirac Hamiltonian (for one half-period):
\begin{equation}
\label{phiDs}
    \phi^D\approx\frac{\pi\alpha\sqrt{\xi}}{4\Delta_D^{3/2}},
\end{equation}

This result can be directly compared with the rotation angle of the classical Foucault pendulum, which reads for 1 day:
\begin{equation}
    \phi_{{\mathrm{day}}}=2\pi-\int \kappa~dS=2\pi\left(1-R^2\int\limits_0^{\theta_0}\kappa(\theta)\sin\theta~d\theta\right)
\end{equation}
where $\theta=\pi/2-\varphi$ is the polar angle corresponding to the latitude $\varphi$. 
The first term is due to the non-inertial nature of the system (the rotation of Earth), whereas the second one is due to the curvature of the Earth's surface. The importance of the Foucault pendulum in popular education is therefore doubled: it allows demonstrating the Earth's rotation and the Earth's curvature, providing two important arguments against the modern "Flat Earth" trend~\cite{mcintyre2021talking}.

This last term is also called the phase of the Foucault pendulum:
\begin{equation}
    \phi_F=\int\kappa~dS
\end{equation}
For a constant curvature of a sphere, one recovers the famous formula
\begin{equation}
    \phi_{\mathrm{day}}=2\pi\cos\theta=2\pi\sin\varphi
\end{equation}
However, if we imagine that the curvature is not constant, but only approximately constant around $\theta=0$, then we obtain an expression similar to the one of the anomalous Hall drift:
\begin{equation}
    \phi_{\mathrm{day}}=2\pi-2\pi R^2\kappa(0)\left(1-\cos\theta\right)
\end{equation}
In the Dirac case, $\phi^D\sim B(0)\times k_{\mathrm{max}}^2$ where $B(0)$ is the Berry curvature maximum. For the Foucault pendulum, the curvature-induced correction (the Foucault phase): $\phi_1\sim \eta(0)\times \theta^2$, where $\eta=\kappa/R^2$ is the deviation of the Earth's curvature $\kappa$ from a sphere.

\subsection{Non-adiabaticity and anharmonicity}

First, we provide the expression for the quantum metric in the Dirac Hamiltonian:
\begin{equation}
    g_{kk}=\frac{\alpha^2}{4\left(\Delta^2+\alpha^2 k^2\right)}.
\end{equation}
The low-wavevector limit of this expression is used for the calculation of non-adiabatic fraction in the main text. We also note that the calculation of anharmonicity for the Dirac case gives exactly the same result for $Q$ as the calculation of non-adiabaticity, because there is no independent mass in the Dirac equation: everything is determined by the interplay of $\Delta$ and $\alpha k$.

We now compare the contributions of non-adiabaticity and anharmonicity for the TE-TM case, where both can potentially be important. The calculation of anharmonicity is presented in the main text. The calculation of non-adiabaticity follows the same lines as the one presented in the main text for the Dirac case.

The quantum metric reads
\begin{equation}
    g_{kk}=\frac{\Delta^2\beta^2 k^2}{\left(\Delta^2+\beta^2 k^4\right)^2}
\end{equation}
which for small wave vectors can be approximated as
\begin{equation}
    g_{kk}\approx \frac{\beta^2 k^2}{\Delta^2}
\end{equation}
The approximation for the non-adiabatic fraction reads
\begin{equation}
\sqrt{f_{NA}}\approx\int\limits_0^{k_{max}}\sqrt{g_{kk}}dk\approx\frac{\beta k_{max}^2}{2\Delta}    
\end{equation}
The corresponding Q-factor (the inverse of the non-adiabatic fraction) reads
\begin{equation}
    Q_1=\frac{4\Delta^2\hbar^4}{\beta^2\xi^2 m^2 x_0^4}
\end{equation}
whereas the $Q$-factor due to anharmonicity, found in the main text, reads
\begin{equation}
    Q_2\approx \frac{\hbar^4 \Delta}{6\beta^2 m^2 \xi x_0^2}
\end{equation}
Their ratio is 
\begin{equation}
    \frac{Q_1}{Q_2}=\frac{12\Delta}{\xi x_0^2/2}
\end{equation}
As we see, it depends on the ratio between the Zeeman splitting and the initial potential energy of the wavepacket in the harmonic oscillator, with an extra factor $12$. The smaller is the initial deviation $x_0$, the larger is $Q_1$ with respect to $Q_2$. For the parameters presented in the main text, the ratio $Q_1/Q_2\gg 1$, which is why we neglect the losses due to the non-adiabaticity in the main text. For larger $x_0$, everything becomes more complicated, but the most important role is played by the critical value of $x_0$ beyond which $Q$ drops very quickly.

\subsection{Limitations of the perturbative treatment}

The condition for the application of the perturbation theory is that the scale of the corrections should be much smaller than the distance between the levels. The rapid decrease of $Q^D$ for large $x_0$ is due to the resonant effect of the perturbation, exceeding this limit. If we write the Hamiltonian as $H=H_{h.o.}+V$, the scale of the perturbation $V$ is determined by the maximal wavevector $k_{max}$. This perturbation leads to a resonant transfer between the states if its energy scale becomes equal to the splitting between the harmonic oscillator states $\hbar\omega$, which can be written as the following condition:
\begin{equation}
    \sqrt{\alpha^2 k_{max}^2+\Delta_D^2}-\left(\Delta_D+\frac{\alpha^2 k_{\mathrm{max}}^2}{2\Delta_D}\right)=\hbar\omega
\end{equation}
where the mass of the particle is $m=\hbar^2\Delta/\alpha^2$, which gives $\hbar\omega=\alpha\sqrt{\xi/\Delta}$. Solving this equation for $k_{\mathrm{max}}$ and converting it to $x_0$ gives $x_{\mathrm{crit}}\approx 64$~$\mu$m (the analytical expressions are quite cumbersome), which is exactly what we see in numerical simulations.

For TE-TM, similar to the case of the Dirac Hamiltonian, a cutoff for $Q$ is with a very high precision determined by the resonant anharmonicity condition $E(k_{\mathrm{max}})-E_{h.o.}(k_{\mathrm{max}})=\hbar\omega$, which reads
\begin{equation}
    \sqrt{\Delta^2+\beta^2 k^4}-\Delta=\hbar\omega
\end{equation}
This constraint leads to the strong decrease of the rotation angles $\phi_{1/2}$ and $\phi$ for large $x_0$. 

We conclude that one should reduce $\Delta$ to maximize the maximal rotation angle, whereas $x_0$ can be kept as large as possible below the critical value, in order to improve the observability of the oscillations.

\subsection{Uncertainty}

The finesse of the oscillator $Q$ is also the measure of the improvement of the precision of the measurement of the Berry curvature $B$. Indeed, the relative uncertainty of the anomalous Hall deviation $\delta(\Delta y)/\Delta y$ determines the uncertainty on the Berry curvature $\delta B/B$. The amplification of the anomalous Hall deviation in the Foucault pendulum configuration increases the denominator of the first fraction by a factor $Q$, thus reducing the uncertainty on the Berry curvature by the same factor. This would mean reducing the $10\%$ uncertainty on the Berry curvature in Ref.~\cite{gianfrate2020measurement} down to $0.1\%$ with $Q=100$ from Fig.~3 of the main text, a significant improvement. However, $Q$ should be maximized not by reducing $x_0$, but rather by reducing $\Delta$. The limit to this improvement is set by $\Delta>\beta/\sigma^2=4\pi^2\beta\sqrt{\xi m}/\hbar$ (wavepacket size should be smaller than the Berry curvature variation scale), around 1~$\mu$eV.

\subsection{Determination of the Berry curvature profile}
In the main text, we show that the Berry-Foucault pendulum allows finding with high precision the Berry curvature at low wave vectors. In this section, we demonstrate that the knowledge of the Berry curvature at low wave vectors  gives a full description of the profile of the Berry curvature for a know Hamiltonian type. 

In the Dirac Hamiltonian, the Berry curvature is given by 
\begin{equation}
    B_z^D\left(k\right)=\frac{\alpha^2 \Delta_D}{2\left(\alpha^2 k^2+\Delta_D^2\right)^{3/2}}
    \label{DirBers}
\end{equation}
and the Berry curvature at low  wave vector is given by $B^0=\alpha^2/2\Delta^2$. It is easy to show that the Berry curvature $B(k)$ can be written as a single-parameter function of $k$, with this single parameter being precisely $B^0$:
\begin{equation}\label{BB0}
    B(k)=\frac{B^0}{\left(1+B^0k^2/2\right)^{3/2}}.
\end{equation}
because its integral is normalized (the Chern number). It means that measuring $B^0$ with a high precision allows one to find the whole distribution of the Berry curvature with the same precision.

In the TE-TM Hamiltonian, the Berry curvature is given by 
\begin{equation}
    B_z^{TE-TM}\left(k\right)=\frac{2\Delta \beta^2 k^2}{\left(\Delta^2+\beta^2 k^4\right)^{3/2}}
    \label{TETMBers}
\end{equation}
and its low wave vector approximation is $B^0(k)=\eta k^2$, $\eta=2\beta^2/\Delta^2$ being the quantity we determine with high precision with the Berry-Foucault pendulum. It is again straightforward to show that the Berry curvature in this case can also be written as a single-parameter function of $k$, with the single parameter $\eta$:
\begin{equation}\label{BB02}
    B(k)=\frac{\eta k^2}{\left(1+\eta k^4/2\right)^{3/2}}.
\end{equation}

Thus, the Berry-Foucault pendulum allows determining the whole Berry curvature distribution in the case of TE-TM as well.

\subsection{Large-amplitude oscillations}

Here, we discuss the case of  $k_{max}\to\infty$. This regime is subject to strong non-adiabaticity due to the possibility of resonant transitions between the harmonic oscillator states. Nevertheless, one can imagine that the maximal value of the AHE drift is achieved for $k$ below this critical condition. Indeed, the maximal AHE drift is 
\begin{equation}
\Delta y^{max}=\sqrt{\frac{\beta}{\Delta}}\frac{\Gamma^2(3/4)}{\sqrt{\pi}}
\end{equation}
It is achieved when $k_{max}\gg k^*= \sqrt{\Delta/\beta}$, which is the wave vector of maximal Berry curvature, which gives $x_0\gg \hbar\sqrt{\Delta}/\sqrt{\beta\xi m}$. For large $\hbar \omega$, this wave vector could  be accessible. The maximal angle of rotation before desynchronization in this limit is 
\begin{equation}
    \phi_{max}^{\infty}=\frac{y_{AHE}^{max}}{x_0}\ll\frac{\beta}{\Delta}\frac{\sqrt{\xi m}}{\hbar}\frac{\Gamma^2(3/4)}{\sqrt{\pi}}
\end{equation}
However, the width of the wave packet in the harmonic oscillator is $l^2=\hbar/\sqrt{\xi m}$. Requiring the expression on the right to be much larger than unity, we obtain $l\ll 2\pi/k^*$, and considering that $l$ determines the size of the wavepacket in the reciprocal space, $\Delta k\gg k^*$, which means that the wave packet has to be so small in real space and so large in reciprocal space, that the semi-classical theory is inapplicable. The limit $k_{max}\to \infty$ is therefore impossible to combine with large rotation angles in a harmonic oscillator in the adiabatic regime, and thus the low-$k$ limit considered above is the most relevant.

\end{document}